\documentclass[usenatbib]{mn2e}
\usepackage{epsfig}
\begin{document}
 
 \title[The Three Faces of $\Omega_m$: Testing Gravity with SN Ia
 Surveys]{The Three Faces of $\Omega_m$: Testing Gravity with Low and
 High Redshift SN Ia Surveys}

\author[Abate and Lahav]{Alexandra Abate$^*$ and  Ofer Lahav$^\dagger$ \\ Department of Physics \& Astronomy, University College London, Gower Street, London, WC1E 6BT, UK}

\maketitle

\begin{abstract}
Peculiar velocities of galaxies
hosting Type Ia supernovae generate a significant systematic effect in
deriving the dark energy equation of state $w$, at level of
a few percent.  Here we illustrate how the peculiar velocity effect in
SN Ia data can be turned from a ``systematic'' into a probe of
cosmological parameters.  We assume a flat $\Lambda$-Cold Dark Matter
model ($w=-1$) and use low and high redshift SN Ia data to derive
simultaneously three distinct estimates of the matter density
$\Omega_m$ which appear in the problem: from the geometry, from the
dynamics and from the shape of the matter power spectrum.  We find
that each of the three $\Omega_m$'s agree with the canonical value
$\Omega_m=0.25$ to within $1\sigma$, for reasonably assumed
fluctuation amplitude and Hubble parameter.  This is consistent with the
standard cosmological scenario for both the geometry and the growth of
structure.  For fixed $\Omega_m = 0.25$ for all three $\Omega_m$'s, 
we constrain
$\gamma = 0.72 \pm 0.21$ in the growth factor $\Omega_m(z)^{\gamma}$,
so we cannot currently distinguish between standard Einstein gravity and predictions from some
modified gravity models.  Future surveys of thousands of SN Ia,
or inclusion of peculiar velocity data, 
could significantly improve the above tests.

\end{abstract}

\section{Introduction}
\renewcommand{\thefootnote}{\fnsymbol{footnote}}
\setcounter{footnote}{1}
\footnotetext{E-mail: aabate@star.ucl.ac.uk}
\setcounter{footnote}{2}
\footnotetext{E-mail: lahav@star.ucl.ac.uk}

\begin{figure}
\center
\epsfig{file=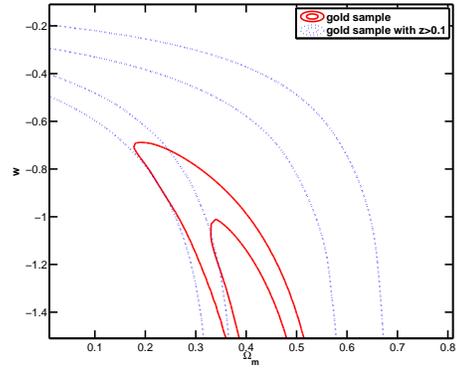,width=6cm,angle=0}
\caption{Assuming a flat universe with a constant equation of state w, these are 1 and 2 $\sigma$ likelihood contours showing the constraints in the $\Omega_m$-$w$ plane for the Riess et al 2007 supernova data.  The red/dark contours are when using the full gold sample (182 SN Ia), the blue/light contours are when using the gold sample but removing all supernovae with redshifts less than 0.1 (146 SN Ia)}
\label{nolowz}
\end{figure}

The observed present acceleration of the universe was first confirmed
a decade ago by two separate groups using Type 1a supernovae
\cite[SN Ia,][]{Perl98, Riess98}.  
SN Ia are one
of a number of probes needed to obtain tighter constraints on dark
energy equation of state, including any possible time evolution.  This
will require surveys of thousands of supernovae out to high redshifts
to accurately measure their luminosity distances from which parameters
describing the dark energy can be inferred.  To achieve the desired
constraints on dark energy, in particular a few percent constraint on
the equation of state parameter $w$, the supernovae will have to be
accurately calibrated.  
It is therefore vital that this calibration is done accurately, and it is the low redshift supernovae which are vital to achieve this, for details see \cite{NSF}.  At low redshift the supernovae distances have little or no dependence on the cosmological parameters such as $\Omega_m$, $\Omega_{\Lambda} $ and the dark energy equation of state $w$.  They do however put a tight constraint on a combination of what is essentially the calibrated magnitude zeropoint ($M$) and the Hubble constant $H_0$, whereas for the high redshift supernovae there is a strong  degeneracy between $M$, $H_0$ and the cosmological parameters of interest.  Figure \ref{nolowz} illustrates the importance of the low redshift supernovae in anchoring the Hubble diagram.  It shows the gold sample from \cite{Riess07} constraints on $\Omega_m$ and $w$ with and without supernovae with redshifts less than $0.1$.   One can see that without the low redshift supernovae (blue/light contours, using 146 SN Ia) the constraints blow up significantly compared to the full gold sample (red/dark contours, using 182 SN Ia).  There are several sources of systematic error which affect the calibration of the zeropoint, for example dust extinction, luminosity evolution, weak lensing,  and Malmquist bias \cite[see][for more details]{Sys}.  This type of error is not decreased by having a large number of supernovae and will necessarily come to dominate the error budget.  The systematic errors mentioned above have long been discussed in the literature and are not considered in this Letter.  There is a source of error which is unique in the fact that it affects only the low redshift ``calibrating" supernovae, their peculiar motions relative to the Hubble flow. 
 
Previous authors have set about using SN Ia to quantify the
degradation of dark energy errors due to peculiar motions or use them
to trace the peculiar velocity field itself in a variety of ways.
Three distinct approaches to this
have been discussed recently in the literature. 
In \cite{NHC07} different flow models based
on the IRAS PSCz survey \citep{branchini99} were used to ``correct" the
luminosity distances by the known peculiar velocities 
before fitting them for the cosmological parameters of
interest. They find the potential systematic error in $w$ caused by
ignoring peculiar velocities is of the order of 4 percent, i.e. quite significant. 

\cite{RSLH} compared peculiar velocities from 98 local supernovae with the gravity field predicted from IRAS. In \cite{danish07} an angular expansion of the radial velocity field
was used to probe the local dipole and quadrupole of the velocity field
at three different distances. 
They found that the dipole is
consistent with galaxy surveys \cite[e.g.][]{erdogdu06} at the same Hubble flow depths.

The third and somewhat different method is utlised by
\cite{Hui06,Cooray06,GLS07} who take a covariance matrix approach.
From the fluctuation in the luminosity distance induced by the
peculiar motions, \cite[see][for derivations]{Hui06, PB04,PB96,
sasaki87}, a covariance matrix for the resulting errors in the
luminosity distance (or similarly the apparent magnitude) can be
calculated.  
The covariance matrix depends on cosmological parameters which
describe the growth and distribution of structure.
In addition to the peculiar velocity effect this is
due to gravitational lensing effect, which is important for redshifts
larger than 1, and we shall ignore it in this Letter.

\cite{Cooray06} found that peculiar
velocities of the low redshift supernovae may prevent measurement of
$w$ to better than 10 percent, and diminish the resolution of the time
derivative of $w$ projected for planned surveys.  \cite{GLS07} used the covariance matrix approach on current data, showing the
changing constraints on $\sigma_8$, $\Omega_m$ and $w$ depending on
the exact redshift range of the SN Ia sample and whether the full
covariance was included or not. They also apply the analysis to
forecasting constraints for future surveys.

Here we unify the analysis of SN Ia data to study simultaneously fits for the expansion of the universe and the growth of structure. There is plenty of discussion on the possibility that the
accelerated expansion of the universe is caused by a modification of
general relativity on large scales \cite[e.g.]
[and references therein]{durrer08,HL07}.
By 
measuring the growth of structure, which directly
effects the observed peculiar velocity field, information is gained to differentiate between the two scenarios.

The rest of the Letter is organised as follows.  In Section \ref{sec:data} we describe the SN Ia sample used in this Letter. In Section \ref{sec:method} we describe the theory underlying SN Ia analysis in cosmology and the effect of the velocity field.

\section{Data}
\label{sec:data}
We analyse both nearby supernovae ($z \le 0.12$) from \cite{JRK07} and high redshift supernovae  ($z \le 0.176$) from a sample compiled by \cite{davis07} which includes data from \cite{Riess07} and \cite{Essence}.  \cite{davis07} combined the data from the two samples by normalising to the low redshift supernovae they had in common. Following \cite{JRK07} 9 supernovae are excluded from the low redshift set, those that are unsuitable due to bad lightcurve fits. This includes supernovae with their first observation more than 20 days after maximum light, those that are hosted in galaxies with excessive extinction ($A^0_V>2.0$ mag) and one outlier (SN1999e), which appears to have an extremely large peculiar velocity. This leaves 124 supernovae from the \cite{JRK07} data set in the redshift range $z \in [0.0023,0.12]$, and median redshift $\bar{z}=0.017$.  The overlapping SN Ia in the two data sets were used to estimate a small normalising offset to the magnitudes from the \cite{davis07} data set (the extra magnitude error is negligibly small). The same procedure was used by \cite{davis07} in normalising the two high redshift data sets. After eliminating duplicated SN Ia, our combined data set has 271 SNe with $z \in [0.0023, 1.76]$, and $\bar{z} = 0.29$. 

\begin{figure}
\center
\epsfig{file=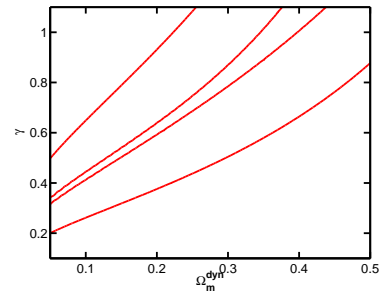, width=5cm}
\caption{The 1 and 2 $\sigma$ contours on $\Omega_m^{dyn}$ and the parameter $\gamma$ from using all 271 low and high redshift SNe. Values of other parameters include $n_s=1$ and $h=0.7$,  $\sigma_v=300$kms$^{-1}$, $\sigma_m=0.1$ and the other two $\Omega_m$'s=0.25.}
\label{fig:gamma}
\end{figure}

\section{Methodology}
\label{sec:method}
We describe here how we utilise the SN Ia dataset described in Section~\ref{sec:data} to estimate cosmological parameters by including the peculiar velocity covariance.

\subsection{Covariance matrix approach}
\label{section:covmatmeth}
The luminosity distance $d_L$ is defined as
\begin{equation}
\mathcal{F}=\frac{\mathcal{L}} {4\pi d_L^2}
\end{equation}
where $\mathcal{F}$ is the observed flux of the supernova and $\mathcal{L}$ is its intrinsic luminosity. The apparent magnitude $m$ of a supernova at redshift $z$ depends on the luminosity distance as follows
\begin{equation}
\label{eq:m}
m(z)=5 \log_{10}D_L(z)-5\log_{10}(H_0)+M+25
\end{equation}
where $M$ is the magnitude zeropoint, and $D_L$ is defined without the Hubble constant as $D_L=H_0d_l$ in kms$^{-1}$.  The equation above ignores the additional terms which involve applying dust corrections, K corrections etc.  For a flat universe containing a matter component and a dark energy component with a constant equation of state, the luminosity distance can be written as
\begin{eqnarray}
\label{eq:dl}
D_L(z)  & = & c(1+z) \int^{z}_0 \frac{dz'}{E(z')} \\
E(z') & = & (\Omega_m(1+z')^3+(1-\Omega_m)(1+z')^{-3(1+w)})^{\frac{1}{2}}
\label{eq:E}
\end{eqnarray}
If the universe was truly homogeneous and isotropic (FRW)
this would be the end of the story, the observed $D_L$ would be described accurately by Eq.~\ref{eq:dl}.  However peculiar velocities have the effect of perturbing the luminosity distance 
\begin{equation}
\label{eq:fracdl}
\frac{\delta D_L}{D_L}=\frac{v_r}{c}\left( 1-\frac{c(1+z)^2}{H(z)d_L(z)}\right)
\end{equation}
where $v_r$ is the radial peculiar velocity of the supernova and
$H(z)$ is the Hubble parameter. See \cite{Hui06, BDG06,PB04,sasaki87} for a derivation.
We emphasize that in the right hand side of the above equation $d_L$ is for an {\it unperturbed} FRW universe, derived at a {\it perturbed} redshift $z$.  

Therefore the covariance of the perturbation in $D_L$, $\delta D_L/D_L$ for a pair $i,j$ is given by
\begin{equation}
\label{eq:covmat}
C_{ij}^L=\frac{\left<v_{ri} v_{rj} \right>}{c^2}\left( 1-\frac{c(1+z)^2}{H(z)d_L(z)}\right)_i\left( 1-\frac{c(1+z)^2}{H(z)d_L(z)}\right)_j
\end{equation}
and
\begin{equation}
\label{eq:si}
\left<v_{ri} v_{rj} \right>=\xi_{ij}=\cos \theta_i\cos \theta_i\Psi_{||}(r)+\sin \theta_i\sin \theta_j\Psi_{\perp}(r)
\end{equation}
is the linear theory radial peculiar velocity correlation function,  \citep{gorski,gjo}. The angles in Eq.~\ref{eq:si} are defined by $\cos\theta_{X}=\hat{\textbf{r}}_{X} \cdot \hat{\textbf{r}}$ and the diagonal elements $\xi_{ii}$ are given by Eq.~\ref{eq:diag} below. The $\Psi_{||}(r)$ and $\Psi_{\perp}(r)$ can be calculated from the matter power spectrum using linear theory
\begin{equation}
\label{eq:psi}
\Psi_{||,\perp}(r)=D'(z_i)D'(z_j)\int \frac{P(k)}{2\pi^2} B_{||,\perp}(kr) dk
\end{equation}
where $P(k)$ is the matter power spectrum, $B_{\perp}=j_0^{'}(x)/x$ and $B_{||}=j_0^{''}(x)$ and $ j_0^{'}, j_0^{''}$ are the first and second derivative of the zeroth order spherical Bessel functions respectively and $D'(z)$ is the derivative of the growth function at redshift $z$. $D'(z)$ is a function of the Hubble parameter and $\Omega_m$. 
See Section \ref{sec:faces} for further discussion. The auto correlation is given by
\begin{equation}
\label{eq:diag}
\xi_{ii}=\frac{1}{3}D'^2(z_i)\int  \frac{P(k)}{2\pi^2}dk.
\end{equation}
We can therefore calculate $C_{ij}^L$ for a pair of supernovae at $z_i$ and $z_j$ respectively given a set of cosmological parameters. 

In the above equations the growth factor is calculated exactly
numerically.  More insight to the dependence on $\Omega_m$ is given by
the commonly used approximation for the growth factor $f = d \ln
\delta / d \ln a \approx \Omega_m(z)^{\gamma}$, where $\gamma \approx
0.6$ \citep{peebles80}, with little dependence on the cosmological
constant \citep{lahav91}, and a slight dependence on $w$ \citep{ws98}.
Recent refined calculations predict $\gamma =0.55$ for the concordance
model, and $\gamma = 0.69$ \citep{LC07} for a particular modified
gravity model, DGP braneworld gravity \citep{DGP}, though this is just an example of many possible modified gravity models.  Below we shall constrain $\gamma$ from the SN Ia data.

\begin{figure}
\epsfig{file=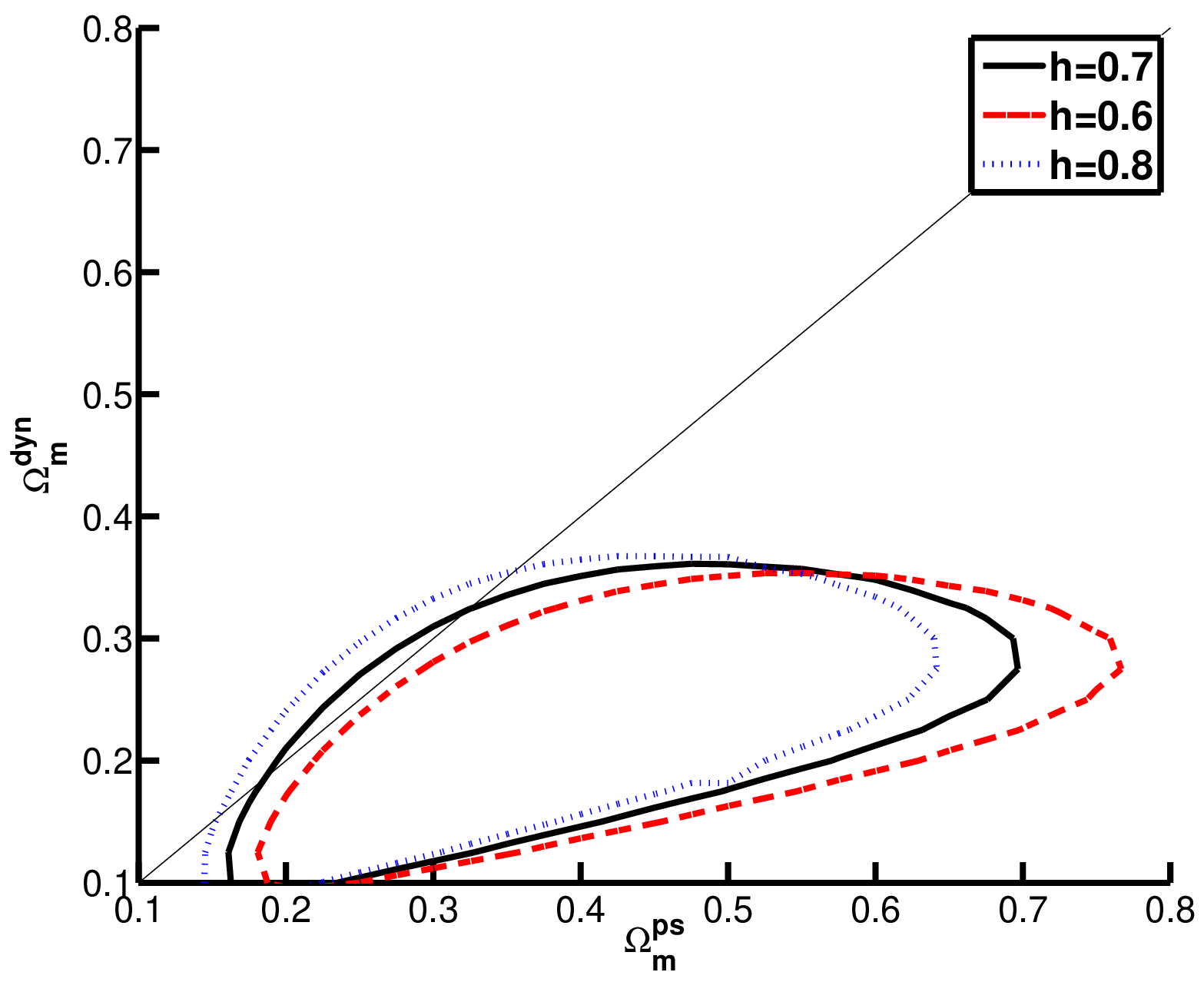,width=5cm,angle=0} 
\epsfig{file=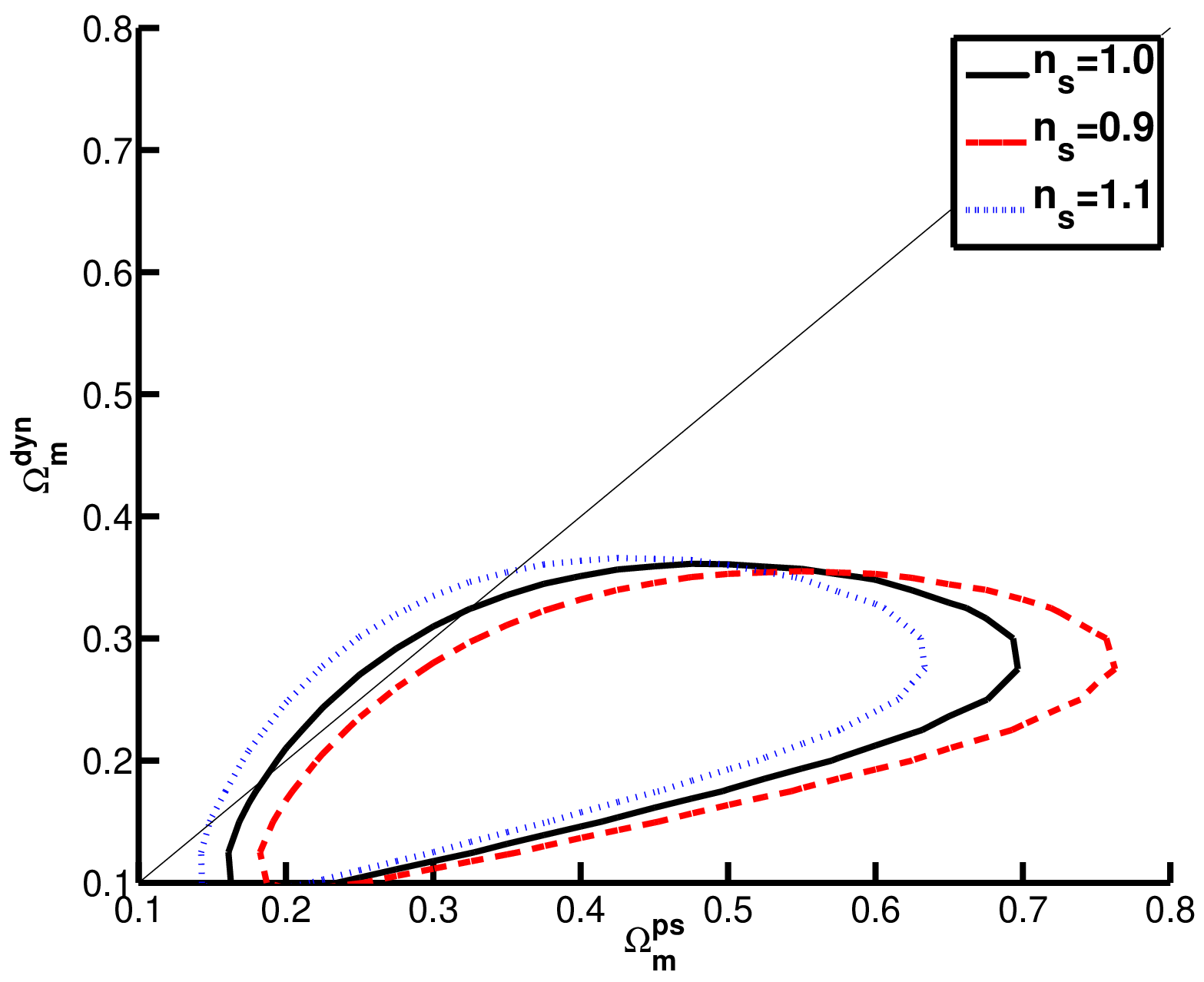,width=5cm,angle=0}
\caption{Constraints on $\Omega_m^{ps}$ and $\Omega_m^{dyn}$ from using all 271 low and high redshift SNe. In both panels $\Omega_m^{geom}$ has been fixed at $0.25$, $\sigma_v$ at $300$kms$^{-1}$ and $\sigma_m$ at $0.1$.  In the top panel $n_s$ has been fixed at $1.0$, and the red (dashed) contours, black (dark) contours and blue (light) contours are the results when $h=0.6$, $h=0.7$ and $h=0.8$ respectively.  In the bottom panel $h$ has been fixed at $0.7$, and the red (dashed) contours, black (dark) contours and blue (light) contours are the results when $n_s=0.9$, $n_s=1$ and $n_s=1.1$ respectively.  The black (dark) contours in both panels are exactly the same. The grey line indicates where $\Omega_m^{ps}=\Omega_m^{dyn}$.}
\label{fig:hns}
\end{figure}

\subsection{Likelihood analysis}

To find the set of cosmological parameters $\bf\Theta_{max}=\left[\theta_1 ... \theta_N \right]$ that best fit the data we find the set that maximise the likelihood function.  Assuming that the data and the observational errors are Gaussian random fields the likelihood function can be written as
\begin{equation}
\label{eq:like}
\mathcal{L}=\frac{1}{\sqrt{(2\pi)^N |\bf{\Sigma}|}} \exp \left( -\frac{1}{2} \sum_{i, j}^{N} D_i\,\, ({\bf \Sigma}^{-1})_{ij}\,\, D_j\right).
\end{equation} where $D_i$ is defined as $D_i=\left(D_L^{obs}-D_L(z)\right)/D_L(z)$ and $\Sigma$ is the covariance matrix including the observational noise.  Following \cite{GLS07} we write this as
\begin{equation}
\Sigma_{ij}=C_{ij}^L+\sigma_i^2\delta_{ij}
\end{equation} 
where $\sigma_i$ is the standard uncorrelated error given by
\begin{equation}
\sigma_i^2=\left(\frac{\ln(10)}{5}\right)^2\left(\sigma_m^2+(\mu^{err}_i)^2\right)+\left( 1-\frac{c(1+z)^2}{H(z)d_L(z)}\right)^2_i\frac{\sigma_v^2}{c^2}
\end{equation}
where $\sigma_v$ is often set to 300kms$^{-1}$ and is included to account
for nonlinear contributions to $\xi_{ij}$ 
\cite[which is derived only in linear theory,][]{silb}, and the velocity of the SN within the host galaxy.
Here $\sigma_m$ is the intrinsic magnitude scatter and  $\mu^{err}$ is the error from the light
curve fitting.

\subsection{The Three Faces of $\Omega_m$}
\label{sec:faces}
From the equations in Section~\ref{section:covmatmeth}  it can easily be seen that $C_{ij}^L$ is a function of $\Omega_m$ through\\
\\
\noindent
(i) $\bf \Omega_m^{geom}$: the geometry of the universe, from $H(z)$ and also $d_L(z)$ in Eq.~\ref{eq:fracdl}. For a flat universe  $\Omega_m^{geom} = 1 -  \Omega_{\Lambda}$, with no dependence on other cosmological parameters. $\Omega_m^{geom}$ is most strongly constrained however by the high redshift SNe through Eqs.~\ref{eq:dl} and \ref{eq:E}.
\\
\\
(ii) $\bf \Omega_m^{dyn}$: the growth of structure in the universe, from $D'(z)$ in Eqs.~\ref{eq:psi} and \ref{eq:diag} or the equivalent growth factor paramaterization $f = [\Omega_m^{dyn}(z)]^{\gamma}$.We note a strong degeneracy 
through the product $\sigma_8\Omega_m(z)^\gamma$ where 
\begin{equation}
\Omega_m(z)=\frac{\Omega_m(z=0)(1+z)^3}{\Omega_m(z=0)(1+z)^3+1-\Omega_m(z=0)}
\end{equation}
for a flat universe, our results can be scaled accordingly.\\ 
\\
(iii) $\bf \Omega_m^{ps}$: the matter power spectrum $P(k)$ in Eqs.~\ref{eq:psi} and \ref{eq:diag}. It is wel known that the shape of the power spectrum depend on the product 
$\Gamma = \Omega_m^{ps} h$, with some degeneracy with e.g. the spectral index $n_s$, $\sigma_8$ and baryon and neutrino mass densities $\Omega_b$ and $\Omega_{\nu}$.\\
Please note that Eq.'s \ref{eq:psi} and \ref{eq:diag} contain all of the low redshift $\Omega_m$ terms.
If the $\Lambda$CDM model of the universe is correct then when varying each of these ``faces" of $\Omega_m$ separately the results should be consistent with each other.  If not this suggests that the $\Lambda$CDM model is inconsistent and the data may favour a model which changes the theory of general relativity on large scales or other dark energy models.

\section{Results}
\label{resultsec}

\begin{figure*}
\begin{tabular}{c|c|c|c}
\epsfig{file=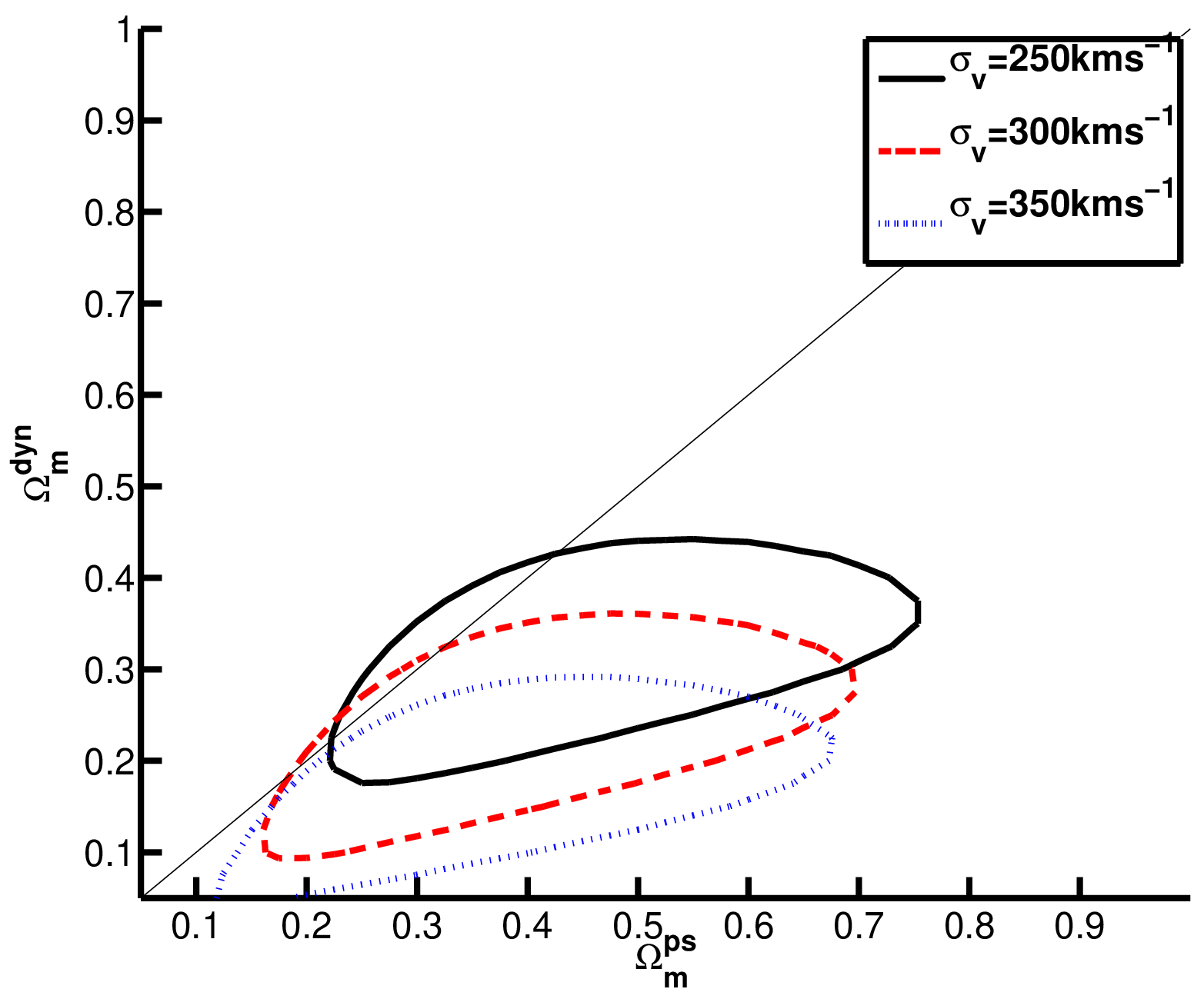,width=4cm,angle=0}& \epsfig{file=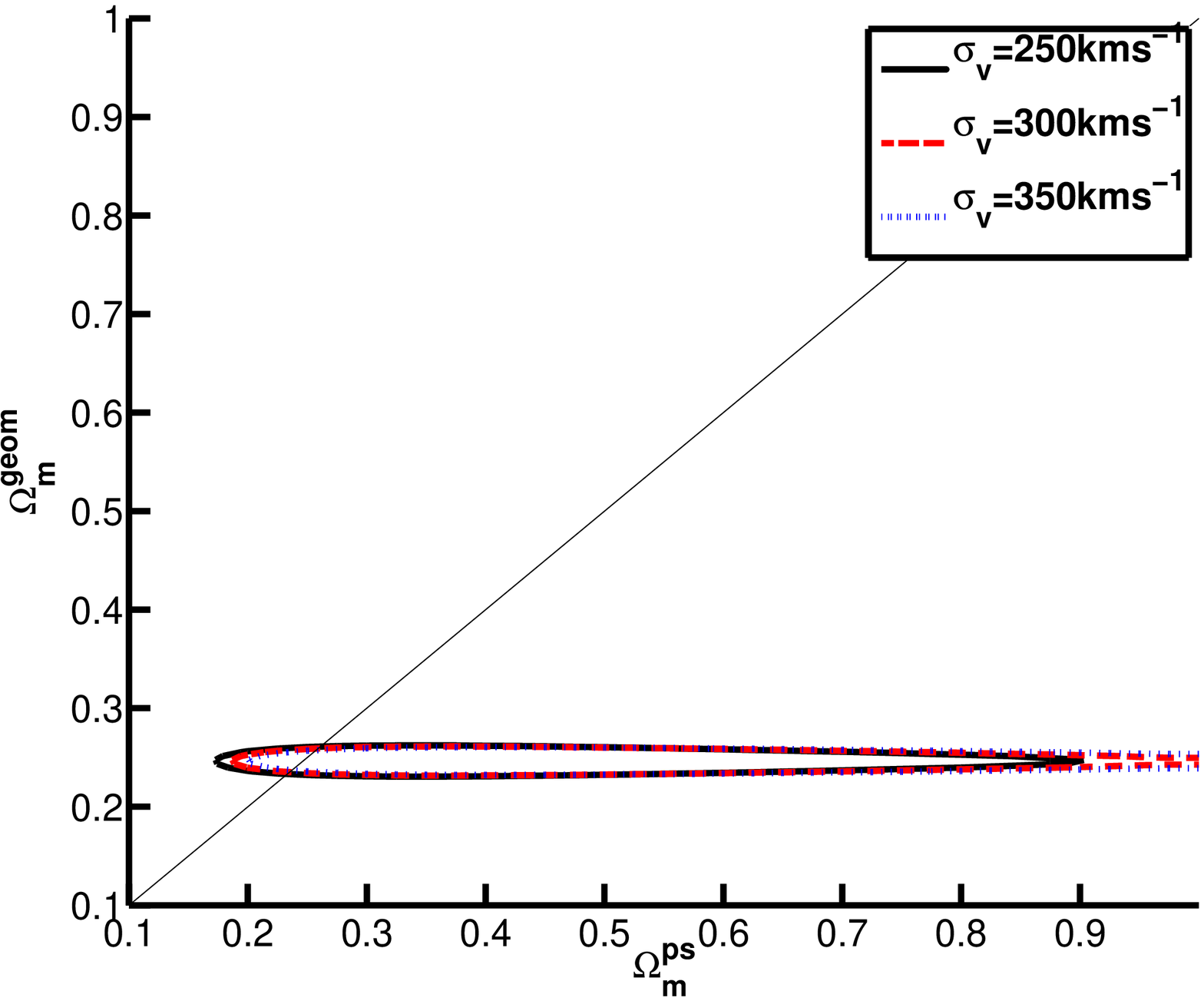,width=4cm,angle=0}&
\epsfig{file=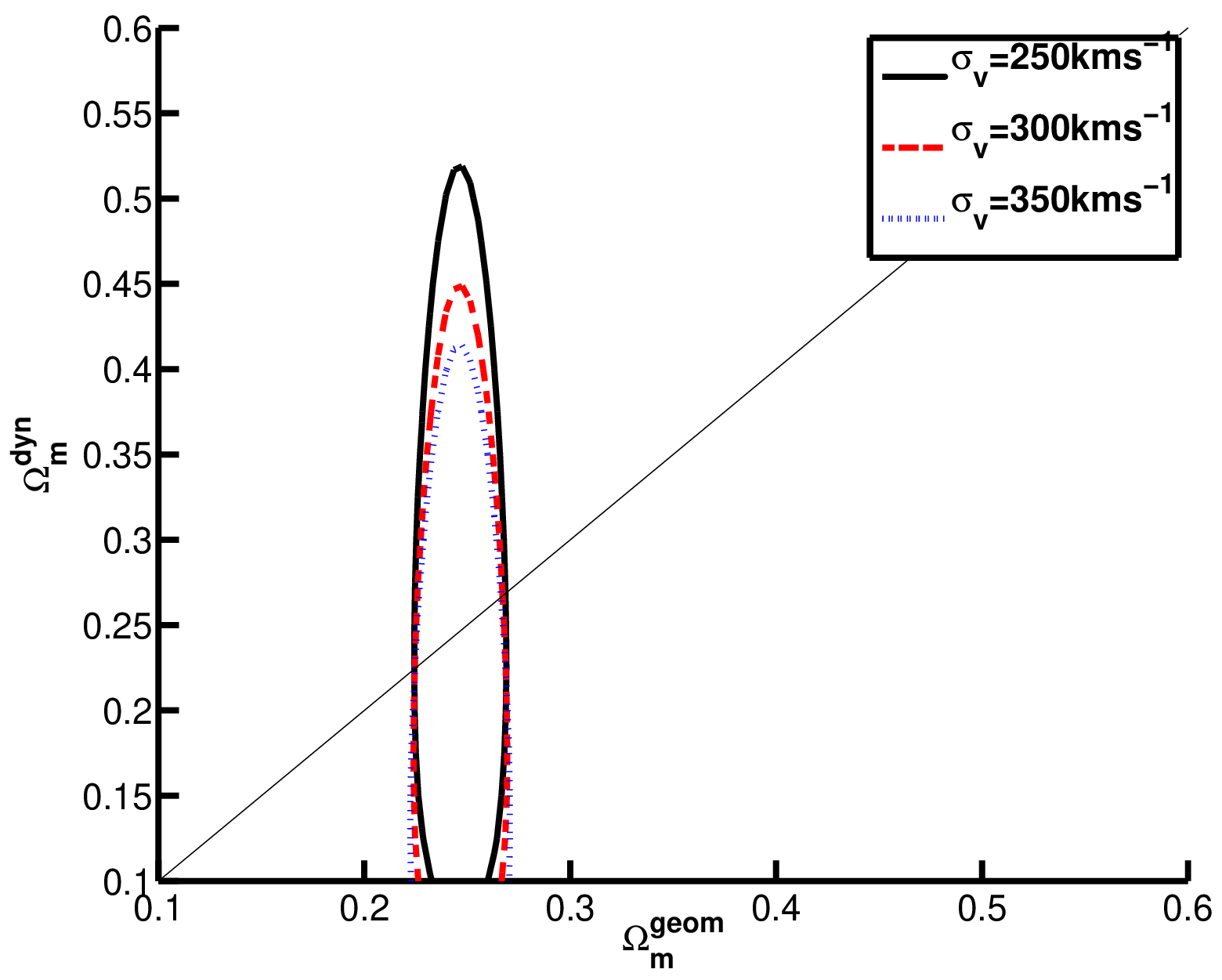,width=4cm,angle=0}&\epsfig{file=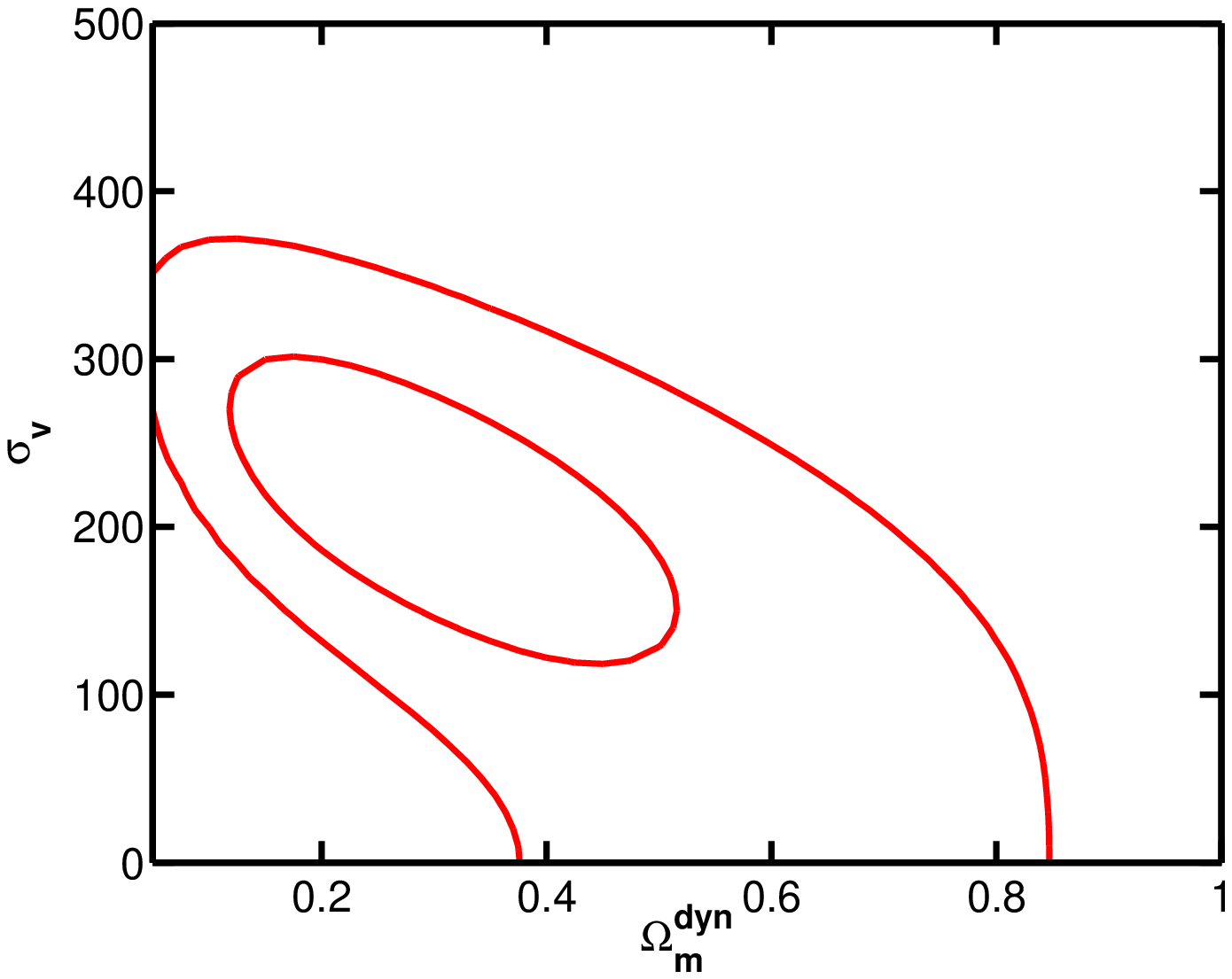,width=4cm,angle=0}
\end{tabular}
\caption{Constraints on all the $\Omega_m$ pairs from using all 271 low and high redshift SNe. From the left the first panel shows $\Omega_m^{ps}$-$\Omega_m^{dyn}$, the second panel $\Omega_m^{ps}$-$\Omega_m^{geom}$, third panel $\Omega_m^{geom}$-$\Omega_m^{dyn}$ and the fourth panel $\Omega_m^{dyn}$-$\sigma_v$.  The black (dark), red (dashed) and blue (light) contours in the first 3 panels correspond to $\sigma_v$ equal to $250$kms$^{-1}$, $300$kms$^{-1}$ and $350$kms$^{-1}$ respectively. Each time $n_s=1$, $h=0.7$, $\sigma_m=0.1$, $\sigma_8=0.8$ and in the first 3 panels the third $\Omega_m$ is kept fixed at $0.25$ and just the 1-$\sigma$ contour is shown.  In the fourth panel $n_s=1$, $h=0.7$, $\sigma_m=0.1$, $\Omega_m^{ps}=\Omega_m^{geom}=0.25$ and both the 1 and 2-$\sigma$ contours are shown. The grey line indicates where $\Omega_m^{x}=\Omega_m^{y}$.}
\label{fig:sv}
\end{figure*}

In the following analysis we assume a flat $\Lambda$CDM universe with
a dark energy equation of state $w=-1$. To gain an insight for the
effect of varying the other cosmological parameters, namely $H_0$, $n_s$, and the ``nuisance" parameters $\sigma_m$ and
$\sigma_v$ we do not marginalise over them but present the results at
some choice values for these parameters. The effect of marginalising over $\sigma_v$ and $\sigma_m$ degrades the error on $\Omega_m^{geom}$ by about 10 percent, the error on $\Omega_m^{ps}$ changes negligibly and the error on $\Omega_m^{dyn}$ degrades by about 30 percent. The larger degradation of the error on $\Omega_m^{dyn}$  is because of its strong degeneracy with $\sigma_v$.
Other parameters are fixed as follows; $\Omega_b=0.04$, $\Omega_\nu=0$ and $\sigma_8=0.8$.  For clarity most contours
have only the 1$\sigma$ confidence level, and where appropriate the
assumed values of other parameters are stated.

\begin{table}
\caption{Results for each ``face" of $\Omega_m$ under different parameter combinations.  Because we do not marginalise the errors are small, but it is still useful to look at the relative errors for the three $\Omega_m$'s. Columns A, C and D set the other two $\Omega_m=0.25$,  whereas column B marginalises over them.  In column A and B the ``nuisance" parameters are $\sigma_v=300$kms$^{-1}$ and $\sigma_m=0.1$.  Column C assumes the ``nuisance" parameters are small ($\sigma_v=200$kms$^{-1}$ and $\sigma_m=0.08$). Column D assumes the ``nuisance" parameters are large ($\sigma_v=400$kms$^{-1}$ and $\sigma_m=0.12$). Each column has set the power spectrum parameters $[h \;\; n_s \;\; \sigma_8]=[0.7 \;\;1.0 \;\;0.8]$. }
\begin{center}
\begin{tabular}{l|l|l|l|l}
\hline
&A&B&C&D\\
$\Omega_m^{geom}$&$0.25_{-0.01}^{+0.01}$&$0.25_{-0.01}^{+0.01}$&$0.25_{-0.01}^{+0.01}$&$0.24_{-0.01}^{+0.01}$\\ \\
$\Omega_m^{dyn}$&$0.25_{-0.14}^{+0.14}$&$0.18_{-0.11}^{+0.11}$&$0.33_{-0.14}^{+0.14}$&$0.10_{-0.14}^{+0.10}$\\ \\
$\Omega_m^{ps}$&$0.43_{-0.26}^{+0.46}$&$0.38_{-0.25}^{+0.45}$&$0.33_{-0.23}^{+0.43}$&$0.48_{-0.26}^{+0.52}$\\
\hline
\end{tabular}
\end{center}
\label{tab}
\end{table}

Table~\ref{tab} shows a set of results for each ``face" of $\Omega_m$ under different parameter combinations.  This table shows that the different  $\Omega_m$'s are consistent with the canonical value of $0.25$ to within 1-$\sigma$. One can see the degeneracy direction of the errors ($\sigma_v$, $\sigma_m$) and $\Omega_m^{dyn}$, discussed below. See table caption for an explanation of the columns. 

Figure~\ref{fig:gamma} shows the constraints on $\Omega_m^{dyn}$ and the parameter $\gamma$, described at the end of Section~\ref{section:covmatmeth}.  
 The contours shown are the 1 and 2 $\sigma$.  For $\Omega_m^{dyn}=0.25$ the 68 percent confidence constraint on $\gamma$ is $\gamma=0.72\pm0.21$ consistent with both Einstein gravity ($\gamma=0.55$) and DGP gravity ($\gamma=0.69$).
 
Figure~\ref{fig:hns} shows the effect of $h$ and $n_s$ on the $\Omega_m^{ps}$-$\Omega_m^{dyn}$ contours.  
As expected there is only a strong influence on $\Omega_m^{ps}$.  One can also see the degeneracy direction of $h$ and $n_s$ in the power spectrum, shown by the indistinguishable differences between the red (dashed) contours in both panels.  
The positions of the red (dashed) contours show that decreasing $h$ by roughly 10 percent is equivalent to increasing $n_s$ also by roughly 10 percent. This is also shown by the blue (light) contours. The contours for $\Omega_m^{ps}$-$\Omega_m^{geom}$ and $\Omega_m^{geom}$-$\Omega_m^{dyn}$ are not shown here since the same combinations of $h$ and $n_s$ as plotted in Figure~\ref{fig:hns} have a nearly negligible effect on the contour positions.  Under different permutations of $h=0.6,0.7,0.8$ and $n_s=0.9,1.0,1.1$ the contours only shift in either the  $\Omega_m^{ps}$ or  $\Omega_m^{dyn}$ direction and even then there is only a maximum shift of $\Delta\Omega_m$ of $0.05$. 

Finally Figure~\ref{fig:sv} shows the effect of $\sigma_v$ on all the contour pairs.  
It has the largest effect on $\Omega_m^{dyn}$ and very little effect on $\Omega_m^{geom}$.  This is because the diagonal elements of $C_{ij}^L$ (see Eq.~\ref{eq:covmat}) are approximately proportional to $(\Omega_m^{dyn})^{2\gamma}$ and to $\sigma_v^2$.  Therefore the best-fit $\Omega_m^{dyn}$ decreases as $\sigma_v$ increases.

\section{Conclusions}
\label{conc}
We have presented in this paper a unified approach
for probing both the expansion of the
universe and the growth of structure with SN Ia data
and to test the consistency
of the $\Lambda$CDM model.
We utilised the SNIa data 
to derive three distinct
estimates of the matter density $\Omega_m$ which appear in the
problem: from the geometry, from the dynamics and from the shape of
the matter power spectrum. We found that each of them agrees with
canonical value $\Omega_m=0.25$ to within 1$\sigma$.  We note we are 
restricting our discussion to $\Lambda$CDM, if we allow $w$ to vary our constraints 
on $\Omega_m^{geom}$ will weaken.
We also constrained $\gamma$ in the growth factor
 $\Omega_m(z)^{\gamma}$ 
and found for $\Omega_m=0.25$, $\gamma=0.72\pm0.21$.
This value of $\gamma$ is consistent with both concordance and some
proposed modified gravity models.

Current and future SN Ia surveys such as SN Factory, GAIA and Skymapper
 (for low redshift), SDSS-II (for intermediate redshift)
and DES, Pan-STARRS, LSST, DUNE and SNAP (for high redshift)
will generate samples of thousands of SN Ia \cite[e.g.][for 
overviews]{Alb06,Pea06}.  Large samples of low redshift SNe will
 greatly improve our 
constraints on 
$\Omega_m^{ps}$ and $\Omega_m^{dyn}$ and the high redshift SNe on 
$\Omega_m^{geom}$.
Utilising 
\textit{galaxy} peculiar velocity data
(using $D_n-\sigma$ and Tully-Fisher distance indicators) 
will also provide improvement on the $\Omega_m^{ps}$ and $\Omega_m^{dyn}$ 
constraints.  Our approach can also be generalised for a range of other
cosmological parameters and exotic models of dark energy and gravity.

\section{Acknowledgments}
We are grateful to Kate Land for providing us with information about
SN Ia data sets, and to Sarah Bridle, Josh Frieman, Chris Gordon, Saurabh Jha, John
Marriner and Jochen Weller for useful conversations.  AA acknowledges
the receipt of a STFC studentship and OL acknowledges a
Royal Society Wolfson Research Merit Award. 
Both AA and OL thank Fermilab and
the Kavli Institute for Cosmological Physics 
at Chicago University for their
hospitality.  We thank the anonymous referee for their detailed comments.



\begin{thebibliography}{99}

\bibitem[\protect\citeauthoryear{Albrecht et al.}{2006}]{Alb06} 
Albrecht A., et al., 2006, astro, arXiv:astro-ph/0609591

\bibitem[\protect\citeauthoryear{Aldering et 
al.}{2002}]{NSF} Aldering G., et al., 2002, SPIE, 4836, 61


\bibitem[\protect\citeauthoryear{Bonvin, Durrer, \& Gasparini}{2006}]{BDG06} 
Bonvin C., Durrer R., Gasparini M.~A., 2006, PhRvD, 73, 023523 

\bibitem[\protect\citeauthoryear{Branchini et al.}{1999}]{branchini99} 
Branchini E., et al., 1999, MNRAS, 308, 1 

\bibitem[\protect\citeauthoryear{Bridle et al.}{2002}]{Bridle02} 
Bridle S.~L., Crittenden R., Melchiorri A., Hobson M.~P., Kneissl R., 
Lasenby A.~N., 2002, MNRAS, 335, 1193

\bibitem[\protect\citeauthoryear{Cooray \& Caldwell}{2006}]{Cooray06} 
Cooray A., Caldwell R.~R., 2006, PhRvD, 73, 103002

\bibitem[\protect\citeauthoryear{Davis et al.}{2007}]{davis07} 
Davis T.~M., et al., 2007, ApJ, 666, 716

\bibitem[\protect\citeauthoryear{Durrer \& Maartens}{2008}]{durrer08}
 Durrer R., Maartens R., 2008, GReGr, 40, 301

\bibitem[\protect\citeauthoryear{Dvali, Gabadadze, \& Porrati}{2000}]{DGP} 
Dvali G., Gabadadze G., Porrati M., 2000, PhLB, 485, 208

\bibitem[\protect\citeauthoryear{Erdo{\u g}du et al.}{2006}]{erdogdu06} 
Erdo{\u g}du P., et al., 2006, MNRAS, 368, 1515


\bibitem[\protect\citeauthoryear{Gordon, Land \& Slosar}{2007}]{GLS07} 
Gordon C., Land K., Slosar A., 2007, PhRvL, 99, 081301

\bibitem[\protect\citeauthoryear{Gorski}{1988}]{gorski} 
Gorski K., 1988, ApJ, 332, L7

\bibitem[\protect\citeauthoryear{Groth, Juszkiewicz, \& Ostriker}{1989}]{gjo} 
Groth E.~J., Juszkiewicz R., Ostriker J.~P., 1989, ApJ, 346, 558

\bibitem[\protect\citeauthoryear{Haugb{\o}lle et al.}{2007}]{danish07}  
Haugb{\o}lle T., Hannestad S., Thomsen B., Fynbo J., Sollerman J., Jha S., 2007, ApJ, 661, 650 

\bibitem[\protect\citeauthoryear{Hui \& Greene}{2006}]{Hui06} 
Hui L., Greene P.~B., 2006, PhRvD, 73, 123526

\bibitem[\protect\citeauthoryear{Huterer \& Linder}{2007}]{HL07} 
Huterer D., Linder E.~V., 2007, PhRvD, 75, 023519 

\bibitem[\protect\citeauthoryear{Kim et al.}{2004}]{Sys} 
Kim A.~G., Linder E.~V., Miquel R., Mostek N., 2004, MNRAS, 347, 909

\bibitem[\protect\citeauthoryear{Jelinsky \& SNAP Collaboration}{2006}]{Snap} 
Jelinsky P., SNAP Collaboration, 2006, AAS, 209, \#98.09

\bibitem[\protect\citeauthoryear{Jha, Riess, \& Kirshner}{2007}]{JRK07} 
Jha S., Riess A.~G., Kirshner R.~P., 2007, ApJ, 659, 122

\bibitem[\protect\citeauthoryear{Kolb et al.}{2005}]{Kolb05} 
Kolb E.~W., Matarrese S., Notari A., Riotto A., 2005, PhRvD, 71, 023524

\bibitem[\protect\citeauthoryear{Lahav et al.}{1991}]{lahav91} 
Lahav O., Lilje P.~B., Primack J.~R., Rees M.~J., 1991, MNRAS, 251, 128

\bibitem[\protect\citeauthoryear{Linder \& Cahn}{2007}]{LC07} 
Linder E.~V., Cahn R.~N., 2007, APh, 28, 481 

\bibitem[\protect\citeauthoryear{Neill, Hudson, \& Conley}{2007}]{NHC07} 
Neill J.~D., Hudson M.~J., Conley A., 2007, ApJ, 661, L123 

\bibitem[\protect\citeauthoryear{Peacock et al.}{2006}]{Pea06} 
Peacock J.~A., Schneider P., Efstathiou G., Ellis J.~R., Leibundgut B., Lilly S.~J., Mellier Y., 2006, astro, arXiv:astro-ph/061090

\bibitem[\protect\citeauthoryear{Peebles}{1980}]{peebles80} 
Peebles P.~J.~E., 1980, The Large Scale Structure of the Universe, Princeton University Press

\bibitem[\protect\citeauthoryear{Peel \& Knox}{2003}]{Peel03} 
Peel A., Knox L., 2003, NuPhS, 124, 83

\bibitem[\protect\citeauthoryear{Perlmutter et al.}{1999}]{Perl98} 
Perlmutter S., et al., 1999, ApJ, 517, 565

\bibitem[\protect\citeauthoryear{Pyne \& Birkinshaw}{1996}]{PB96} 
Pyne T., Birkinshaw M., 1996, ApJ, 458, 46 

\bibitem[\protect\citeauthoryear{Pyne \& Birkinshaw}{2004}]{PB04} 
Pyne T., Birkinshaw M., 2004, MNRAS, 348, 581

\bibitem[\protect\citeauthoryear{Radburn-Smith, Lucey, \& Hudson}{2004}]{RSLH} Radburn-Smith D.~J., Lucey J.~R., Hudson M.~J., 2004, MNRAS, 355, 1378

\bibitem[\protect\citeauthoryear{Riess et al.}{1998}]{Riess98} 
Riess A.~G., et al., 1998, AJ, 116, 1009 

\bibitem[\protect\citeauthoryear{Riess et al.}{2007}]{Riess07} 
Riess A.~G., et al., 2007, ApJ, 659, 98

\bibitem[\protect\citeauthoryear{Sasaki}{1987}]{sasaki87} 
Sasaki M., 1987, MNRAS, 228, 653

\bibitem[\protect\citeauthoryear{Silberman et al.}{2001}]{silb} 
Silberman L., Dekel A., Eldar A., Zehavi I., 2001, ApJ, 557, 102

\bibitem[\protect\citeauthoryear{Sugiura, Sugiyama, \& Sasaki}{1999}]{sss99} 
Sugiura N., Sugiyama N., Sasaki M., 1999, PThPh, 101, 903 

\bibitem[\protect\citeauthoryear{Wang \& Steinhardt}{1998}]{ws98} 
Wang L., Steinhardt P.~J., 1998, ApJ, 508, 483 

\bibitem[\protect\citeauthoryear{Wood-Vasey et al.}{2002}]{nsf} 
Wood-Vasey W.~M., et al., 2002, AAS, 34, 1205 

\bibitem[\protect\citeauthoryear{Wood-Vasey et al.}{2007}]{Essence} 
Wood-Vasey W.~M., et al., 2007, ApJ, 666, 694 

\end{thebibliography}
\end{document}